\pdfoutput=1

\documentclass[aps, prl, amsmath, amssymb, showpacs, superscriptaddress, twocolumn]{revtex4-1}

\usepackage{verbatim,bbm,graphicx,color}

\graphicspath{{Figures/}}

\newcommand{\bra}[1]{\ensuremath{\langle #1 |}}
\newcommand{\ket}[1]{\ensuremath{| #1 \rangle}}
\newcommand{\be}{\begin{equation}}
\newcommand{\ee}{\end{equation}}
\newcommand{\ie}{{\it i.e.}\ }
\newcommand{\eg}{{\it e.g.}\ }

\begin{document}

\title{Incoherent Control of the Retinal Isomerization in Rhodopsin}
\author{Felix Lucas}
\affiliation{Max Planck Institute for the Physics of Complex Systems, N\" othnitzer Stra\ss{}e 38, 01187 Dresden, Germany}
\affiliation{University of Duisburg-Essen, Faculty of Physics, Lotharstra\ss{}e 1-21, 47057 Duisburg, Germany}
\author{Klaus Hornberger}
\affiliation{University of Duisburg-Essen, Faculty of Physics, Lotharstra\ss{}e 1-21, 47057 Duisburg, Germany}

\date{\today}

\begin{abstract}
We propose to control the retinal photoisomerization yield through the back-action dynamics imparted by a nonselective optical measurement of the molecular electronic state. This incoherent effect is easier to implement than comparable coherent pulse shaping techniques, and is also robust to environmental noise. A numerical simulation of the quantum dynamics shows that the isomerization yield of this important biomolecule can be substantially increased above the natural limit.
\end{abstract}

\maketitle

\emph{Introduction.---}\emph{Coherent control} schemes steer a quantum system towards a specific target state by designing constructive or destructive interferences between different pathways through external laser fields  \cite{Rabitz1993, Rabitz2000}. This concept, which exploits the superposition principle in isolated quantum systems, was first used to improve conventional photochemical methods, for example the selective making and breaking of chemical bonds in molecules \cite{Zewail1988a, Zewail1988b, Shapiro2003}. Later on, it proved to be useful in other areas, such as solid state physics {\cite{Lukin2006}} or quantum information {\cite{Viola1998}}.

For the sake of concreteness, we focus on a prototypical problem of biomolecular physics: how to control the photoisomerization dynamics of retinal, \ie the torsional motion from the molecular trans- to the cis-configuration, taking place within the visual pigment protein rhodopsin. This isomerization reaction constitutes the primary step in human vision. It was shown to take place in the short period of 200 fs \cite{Shank1991, Shank1993a} and to result in isomerization yields as high as 65\% \cite{Dartnall1968, Mathies2001, Mathies2010}, likely due to evolutionary optimization.

The task of manipulating such biomolecules under natural conditions is extremely challenging for current coherent control techniques \cite{Motzkus2002, Miller2006, Rabitz2009, Altucci2012}. This is due to the multitude of densely spaced electronic and ro-vibrational energy levels, making it difficult to address single specific states and leading to a rapid redistribution of the excited state population to unwanted degrees of freedom. The unavoidable environmental noise and decoherence degrades the quantum efficiency further and limits the accuracy of the control process \cite{Khasin2011}.

To avoid these limitations of coherent control, we seek to exploit the \emph{incoherent} dynamics encountered in open quantum systems. Control schemes based on manipulating steady state properties of an open system \cite{Zoller1996, Knight1999, Parkins2003, Parkins2006, Morigi2007, Zoller2008, Viola2009, Cirac2009, Schirmer2010, Soerensen2011} and applying measurement-conditioned state transformations \cite{Jacobs1999, Wiseman2001, Roa2006, Moelmer2007, Wiseman2010} have demonstrated the potential and intrinsic robustness of incoherent control schemes: Unlike in a coherent evolution, incoherent dynamics may be designed to yield a specific final state independently of  the initial and intermediate system states and of possible environmental noise. The mentioned approaches are, however, realistic only in highly engineered systems \cite{Blatt2011, Polzik2011, Wineland2013, Haroche2011, Siddiqi2012, Riste2012, Meschede2012, Hanson2014}, since they require tuning a complex, in practice inaccessible molecular environment or detecting the outcomes of delicate quantum measurements.

A different incoherent control approach, which is suitable for a wide range of experimental scenarios, uses the measurement back-action associated with nonselective measurements to guide a wave packet, and to suppress unwanted transitions \cite{Mendes2003, Rabitz2006a}. The read-out is not recorded in such a setting \cite{Breuer2002, Wiseman2010}, so that any process can be used that would \emph{in principle} allow an observer  to distinguish between specific system states. For instance, if the states differ  appreciably in their photon scattering or absorption cross section, the radiation field will acquire time-resolved information about the molecule by illuminating it at suitable points in time. This requires a pulsed laser source, but no advanced pulse shaping techniques, illustrating how complex systems in an ambient environment can be influenced even in the typical situation that only a limited number of handles is available.

In this letter, we propose a realistic control scenario of a complex molecule under ambient conditions in which measurement-based incoherent control clearly outperforms its coherent control counterpart in terms of efficiency and robustness. Specifically, we propose to steer the configuration state wave packet of retinal by inducing controlled decoherence between its ground and first excited electronic state. The different infrared absorption spectra \cite{Kukura2005} allow probing the electronic state via vibrational or two-photon Raman transitions \cite{Diller2002, Kukura2005}. This way, the isomerization yield can be enhanced substantially, surpassing established coherent pulse shaping techniques \cite{Miller2006}.

We start by introducing a model Hamiltonian, which describes the isomerization dynamics of retinal involving two electronic states. We then set up the Markovian master equation that accounts for the decoherence induced by the continuous nonselective measurement effected by the laser interaction. After outlining the experimental implementation, we discuss the dynamics in presence of controlled decoherence, and determine the isomerization yield as a function of the measurement rate.

\begin{figure}[t]
  \centering
  \includegraphics[width=\columnwidth]{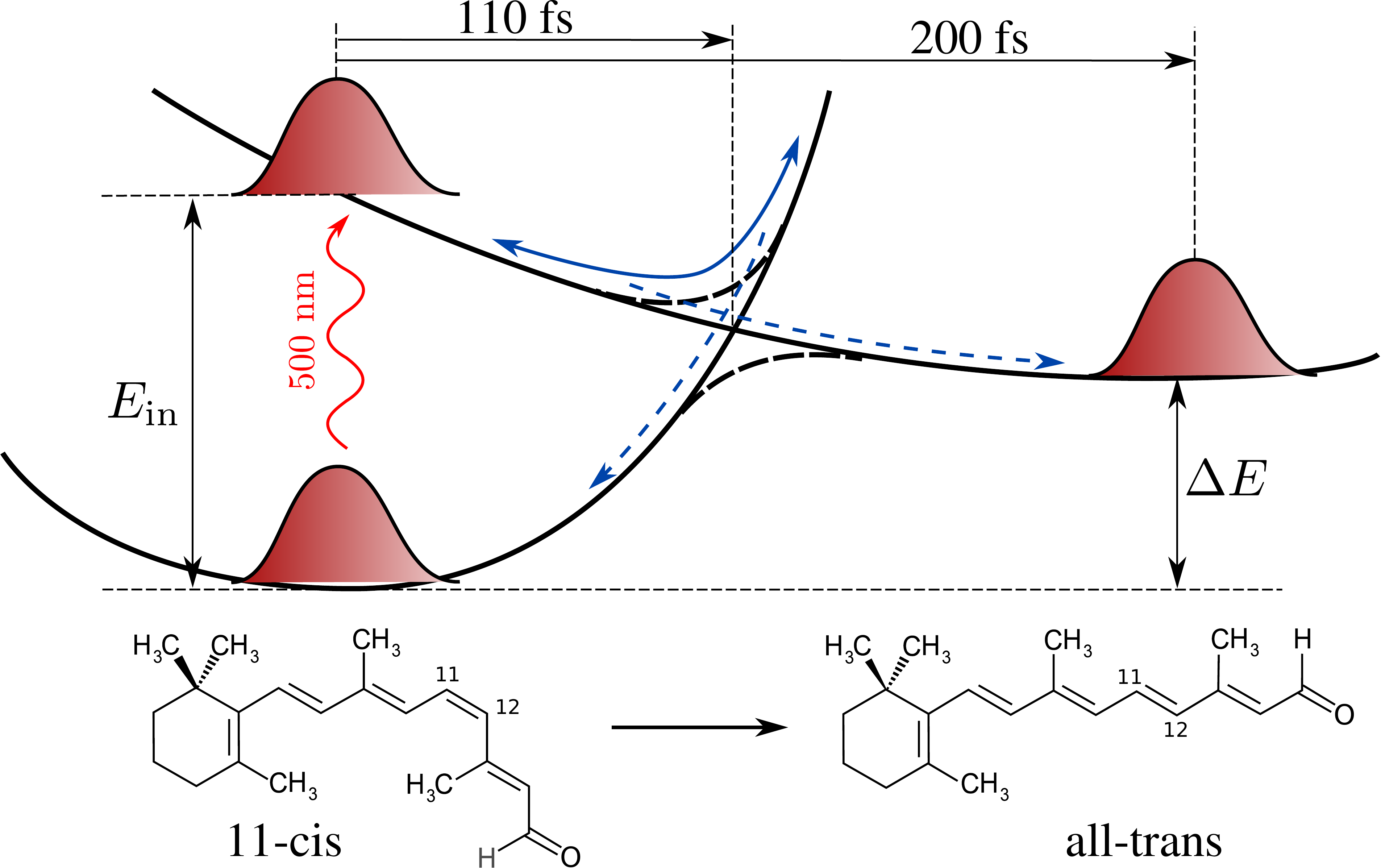}
  \caption{Schematic of the retinal isomerization dynamics. A $500 \,\text{nm}$ photon promotes the ground state of the 11-cis configuration to the first excited electronic potential. After up to seven partial transits through the anti-crossing with the ground state potential, about 65\% of the total population is found in the all-trans state \cite{Mathies2010}. The excess energy at the potential minima gets dissipated rapidly to other vibrational degrees of freedom.
\label{fig:retinalpotentials}}
\end{figure}

\emph{Two state isomerization model.---}The photoisomerization of retinal in rhodopsin is essentially a rotation about the $\mathrm{C}_{11}$=$\mathrm{C}_{12}$ double bond \cite{Wald1968}, as indicated by the structural formula in Fig.~\ref{fig:retinalpotentials}. Starting out in the minimum of the lowest lying electronic potential, which corresponds to the 11-cis isomer, the molecule is excited to the first electronic potential by absorbing a $500 \,\text{nm}$ photon. The excited state wave packet then evolves along the isomerization coordinate, reaching an avoided crossing with the Born-Oppenheimer surface of the electronic ground state after about $110 \,\text{fs}$ {\cite{Mathies2010}}, see Fig.~\ref{fig:retinalpotentials}. Part of the population stays in the excited state and proceeds within about $200 \,\text{fs}$ towards a potential minimum corresponding to a highly twisted all-trans photoproduct, which then relaxes into the vibrational ground state of the all-trans isomer within about $40\,\text{ps}$ \ cite{Shank1991,Shank1993a}, with near unit efficiency. The other part tunnels to the electronic ground state and, after getting reflected further uphill, returns to the avoided crossing where partial tunneling occurs again. This sequential tunneling continues in a coherent fashion for about 1 to $2\,\text{ps}$ \cite{Shank1994}. In the end, about 65\% of the total population is found in the all-trans retinal isomer \cite{Mathies2001}.

A simple, but sufficiently realistic model of the described time evolution is given by  two coupled harmonic potential energy surfaces with frequencies $\omega_1$ and $\omega_2$,
\begin{align}
 \mathsf{H}_\text{ret} &= \frac{\hbar^2 \mathsf{k}^2}{2 m} \otimes \mathbbm{1} + \frac{m}{2} \omega_1^2  \mathsf{x}^2 \otimes | 1 \rangle \langle 1| + \Big[ \Delta E + \frac{m}{2} \omega_2^2  \nonumber \\ 
  &\mathrel{\phantom{=}} \times\left( \mathsf{x} - \Delta x \right)^2 \Big] \otimes | 2 \rangle \langle 2|  + \mathbbm{1} \otimes \alpha \hbar \Big(| 1 \rangle \langle 2| + | 2 \rangle \langle 1|\Big). 
  \label{isoschroeq}
\end{align}
Here, $| 1 \rangle$ and $| 2 \rangle$ are the ground and first excited electronic states, $\mathsf{x}$ and $\mathsf{k}$ are the isomerization coordinate and momentum operators, and $\Delta E$ and $\Delta x$ denote the energy and isomerization coordinate offset between the electronic potential curves, see Fig.~\ref{fig:retinalpotentials}. In agreement with experimental data \cite{Shank1991, Shank1993a, Mathies2010}, we have $E_{\text{in}} = hc / 500 \,\text{nm}$ and $\Delta E = 0.6 E_{\text{in}}$.

The isomerization dynamics starts with the initial state $| \psi_0 \rangle$, given by the vibrational ground state of the 11-cis configuration promoted to the excited electronic potential surface, 
\begin{equation}
  | \psi_0 \rangle = \left( \frac{m \omega_1}{\pi \hbar} \right)^{\frac{1}{4}} \int \mathrm{d}x \exp \left[ - \frac{m \omega_1}{2 \hbar} {x}^2 \right]  |x\rangle |2\rangle . \label{vibground}
\end{equation}
Since the mass in (\ref{isoschroeq}) and (\ref{vibground}) only rescales the isomerization coordinate, the remaining parameters are the frequencies $\omega_1$ and $\omega_2$ and the coupling strength $\alpha$. The experimentally determined isomerization timescales are reproduced by $\omega_1 = 2\pi / 300 \,\text{fs}$ and $\omega_2 =2\pi/ 600 \,\text{fs}$, while the observed isomerization yield of 65\% is obtained for  $\alpha = 0.1 \,\text{fs}^{- 1}$. In the numerical simulation we make the physically plausible assumption that wave packets reaching one of the potential minima lose their excess kinetic energy to other vibrational degrees of freedom and thus do not return to the avoided crossing.  This can be implemented by an absorbing imaginary potential located at the bottom of each well.

This model takes into account that the decisive part of the isomerization dynamics consists essentially of a series of separate Landau-Zener tunnelings, as was conjectured based on experimental data \cite{Shank1993b}.  The probability for staying on the diabatic potential surface in a single transit of the avoided crossing is therefore well described by the Landau-Zener probability $\exp[-2\pi\delta]$. With the wave packet velocity at the avoided crossing fixed by the parameters of the two potential energy surfaces, we have the adiabaticity parameter $\delta = 11.5\, \alpha^2 \,\text{fs}^2$. Taking into account that there is no return to the avoided crossing once a potential minimum is reached, the probability for obtaining the all-trans isomer after an odd number $n$ of transits is given by
\begin{equation}
  P_n (\delta) = \exp [- 2 \pi \delta] \sum_{i = 0}^{(n-1)/2}\!\!  (1 - \exp [- 2 \pi \delta])^{2i}. \label{isolz}
\end{equation}
This prediction is well confirmed by the numerically exact solution of the Schr\"odinger equation with (\ref{isoschroeq}). It determines that after seven transits (four from left to right and three from right to left) the isomerization dynamics is complete.  That is, essentially all population is found in one of the potential minima after about $1.1\,\text{ps}$, a period well within the vibrational coherence timescale of retinal (1-2\,ps) \cite{Shank1994}.

\emph{Continuous nonselective measurement of retinal.---}A general quantum measurement is described by measurement operators $\mathsf{M}_i$ (with $\sum_i \mathsf {M}_i^\dag \mathsf {M}_i = \mathbbm {1}$), each associated with a different measurement outcome $i$ \cite{Breuer2002}. The detection of $i$ occurs with probability $p_i=\mathrm{Tr}( \mathsf {M}_i^\dag \mathsf {M}_i\rho)$, transforming the state to $\rho^{(i)} = \mathsf M_i \rho \mathsf M_i^\dag/p_i$. By modeling a \emph{continuous nonselective measurement} as a Poissonian process with rate $\gamma$ and ignoring the outcomes, it follows that the system evolution is given by the Markovian master equation
\begin{equation}
 \partial_t \rho =  \frac{1}{i\hbar} [\mathsf H_\text{ret}, \rho] + \gamma \sum_i\! \left\{\! \mathsf M_i \rho \mathsf M_i^\dag - \frac 1 2  \mathsf M_i^\dag \mathsf M_i \rho - \frac 1 2 \rho \mathsf M_i^\dag \mathsf M_i \!\right\}.
 \label{isons}
\end{equation}

To get a handle on the isomerization of retinal, consider a measurement of its electronic excitation state, as characterized by the projective measurement operators $\mathsf M_1 = \mathbbm 1 \otimes \ket 1 \bra 1$ and $\mathsf M_2 = \mathbbm 1 \otimes \ket 2 \bra 2$. Frequent nonselective measurements then induce a dephasing dynamics in the electronic subspace, which is described by the incoherent part of Eq.~\eqref{isons}.

The electronic state of retinal can be probed optically since the ground and first excited state give rise to strongly different infrared absorption spectra. This is apparent in the time-resolved resonance Raman studies of Ref.~\cite{Kukura2005}, which report the emergence of a distinct infrared signature of the excited state within femtoseconds after the optical excitation---much faster than the picosecond timescale required for completing the isomeric torsion. Specifically, the first excited electronic state exhibits three pronounced peaks in the absorption spectrum between 800 and $950\,\text{cm}^{-1}$, whereas the ground-state absorption cross section  practically vanishes at those energies. This spectral region is shaped by the concerted hydrogen-out-of-plane (HOOP) wagging motion. In particular, the mentioned peaks are associated with the $\mathrm C_{10}$--$\mathrm H$, $\mathrm C_{11}$--$\mathrm H$, and $\mathrm C_{12}$--$\mathrm H$ wagging modes.

The HOOP-mode absorption can be probed experimentally either by directly shining in photons at the mentioned infrared wavelength \cite{Diller2002} or by exciting an optical two-photon Raman transition \cite{Kukura2005}. An absorption event, if detected, would allow one  to infer immediately that the retinal molecule is in its first excited state. Assuming that we cannot detect individual absorption events, a continuous illumination with infrared or Raman photons effects the dynamics described by Eq.~\eqref{isons} with a single Lindblad operator $\mathsf M_2$. Using the completeness relation for the measurement operators, $\mathsf M_2 = \mathbbm{1} - \mathsf M_1$, and the fact that they are both projectors, one obtains the master equation of a complete nonselective continuous measurement, Eq.~\eqref{isons}  with $\mathsf M_1$ \emph{and} $\mathsf M_2$. The associated measurement rate $\gamma$ is then given by half the excited state absorption rate, \ie the impinging photon current at the relevant infrared frequencies multiplied with the associated absorption cross section. Other properties of the laser light, \eg its phase or spectral shape, have no influence while additional sources of dephasing, \eg due to electron-vibrational interactions, even enhance the effect of the nonselective measurement.

\emph{Incoherent control of the retinal iso\-meri\-zation.---}Be\-fore we turn to the numerical solution of the master equation (\ref{isons}) it is instructive to discuss what to expect qualitatively. It is natural to assume that the final isomerization yield will be determined by the individual transits of the avoided crossing. As mentioned above, in the absence of nonselective measurements they are well described by consecutive Landau-Zener tunnelings, see Eq.~\eqref{isolz}. On the other hand, the presence of dephasing, as given by the incoherent part of (\ref{isons}), is known to decrease the Landau-Zener tunneling probability monotonically, from the coherent value $1 - \exp[-2\pi\delta]$ (for $\gamma = 0$) to the lower limit $(1 - \exp [- 4 \pi \delta]) / 2$ in case of infinitely strong dephasing $\gamma \to \infty$. An expression interpolating between the two limits was found based on a highly convergent expansion of the open Landau-Zener dynamics \cite{Lucas2013,Lucas2014}. 

The Landau-Zener analogy suggests to enhance the isomerization yield by applying the nonselective measurement dynamics \eqref{isons} whenever the excited state wave packet transits the avoided crossing from left to right. Of course, one should keep in mind that the Landau-Zener problem describes an idealized avoided crossing of two states, whose unbounded energy difference varies linearly with time. Our isomerization model, in contrast, involves a wave packet in the continuous configuration coordinate $x$ experiencing at most a finite potential energy difference.

\begin{figure}[tf]
  \centering
  \includegraphics[width=\columnwidth]{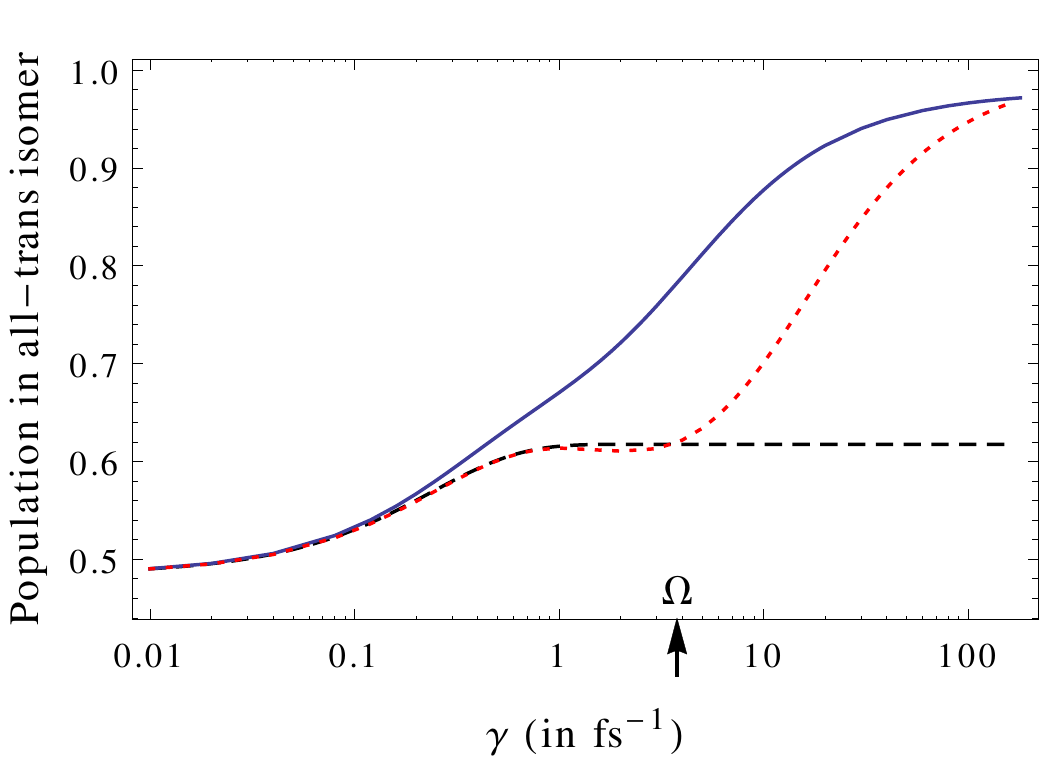}
  \caption{(Color online) Population of the all-trans isomer after the first $200 \,\text{fs}$ as a function of the measurement rate $\gamma$. The red dotted line corresponds to \emph{continuous} nonselective measurement dynamics over the entire $200 \,\text{fs}$ time interval, the black dashed line depicts the corresponding prediction of the analytical open Landau-Zener model \cite{Lucas2013}. One observes that the isomerization yield increases strongly once $\gamma$ exceeds the characteristic coherent system timescale $\Omega$. The blue solid line represents the result of a \emph{pulsed} nonselective measurement, applied only during the time period $t \in [90 \,\text{fs}, 120 \,\text{fs}]$  when the wave packet passes the avoided crossing. Note that the isomerization yield gets enhanced substantially in the pulsed case, even at moderate measurement rates $\gamma\simeq\Omega$.
  \label{fig:popg1}}
\end{figure}

Let us first discuss how a single transit of the avoided crossing in retinal depends on the measurement rate $\gamma$. The red dotted line in Fig.~\ref{fig:popg1} shows the resulting all-trans population as a function of $\gamma$ on a semi-logarithmic scale; it is obtained by numerically propagating the master equation (\ref{isons}) until about $200 \,\text{fs}$. One observes that a high measurement rate of 100\,fs$^{-1}$ would push the single-transit population transfer from below 50\% to around 96\%. This should be compared to the dashed line, which represents the analytical prediction of the Landau-Zener model under dephasing from Ref.~\cite{Lucas2013}. The two agree very well for measurement rates smaller than $2 \,\text{fs}^{- 1}$, but for $\gamma > \Omega\simeq 3.8 \, \text{fs}^{- 1}$ the numerical all-trans population starts rising above the Landau-Zener threshold of $(1  + \exp [- 4 \pi \delta]) / 2\simeq 62\%$, eventually approaching unity. $\Omega$ is the greatest characteristic frequency of the system dynamics, the Rabi frequency immediately after the initial excitation pulse, $\Omega = \sqrt{ (E_\text{in}/\hbar)^2+\alpha^2}$. Such a minimal coherent timescale is absent in the the Landau-Zener model, where the level splitting grows above all bounds. The discrepancy between the Landau-Zener prediction and the more realistic two state model (\ref{isons}) thus illustrates that it is important to account for the finite energy differences in the isomerization dynamics.

An intuitive explanation why the passage through the avoided crossing increases with growing rate $\gamma$ is given by the quantum Zeno effect: Frequent measurements of the electronic state prevent the system from evolving away from an eigenstate of the uncoupled Hamiltonian, which in turn enhances the diabatic transition. In practice, the measurement rate cannot be increased arbitrarily, since retinal can absorb only a finite amount of infrared photons  without disintegrating. The maximal sustainable infrared  power is determined both by the number of absorbing modes (in our case three HOOP transitions)  and by the timescale on which the excess vibrational energy is redistributed to other modes. Although the $\mathrm{C}$--$\mathrm{H}$ bond energy is more than an order of magnitude higher than the energy of a HOOP frequency photon, a rate of $\gamma = 100 \,\text{fs}^{- 1}$ applied over $200 \,\text{fs}$ (corresponding to the right hand side of Fig.~\ref{fig:popg1}) would be too much.

But the vibrational energy absorbed by retinal can be reduced substantially by  switching on the measurement only during the actual transit of the avoided crossing. The population transfer resulting from measurements at times $t \in [90 \,\text{fs}, 120 \,\text{fs}]$ is shown by the solid line in Fig.~\ref{fig:popg1}. The required femtosecond infrared or Raman pulses are feasible with present day technology {\cite{Bromberg1997, Asplund2000, Fecko2003, Fayer2001, Diller2002, Kukura2005}}. This pulsed measurement not only reduces the heating of the retinal molecule by 85\%, but also leads to a substantially increased isomerization yield already at moderate rates. This further increase of the diabatic transition probability results from the fact that besides suppressing Landau-Zener tunneling, dephasing stretches the characteristic tunneling interval \cite{Lucas2013}. Hence, in the case of pulsed dephasing we limit the suppressed, incoherent tunneling to a small time period.

\begin{figure}[t]
  \centering
  \includegraphics[width=\columnwidth]{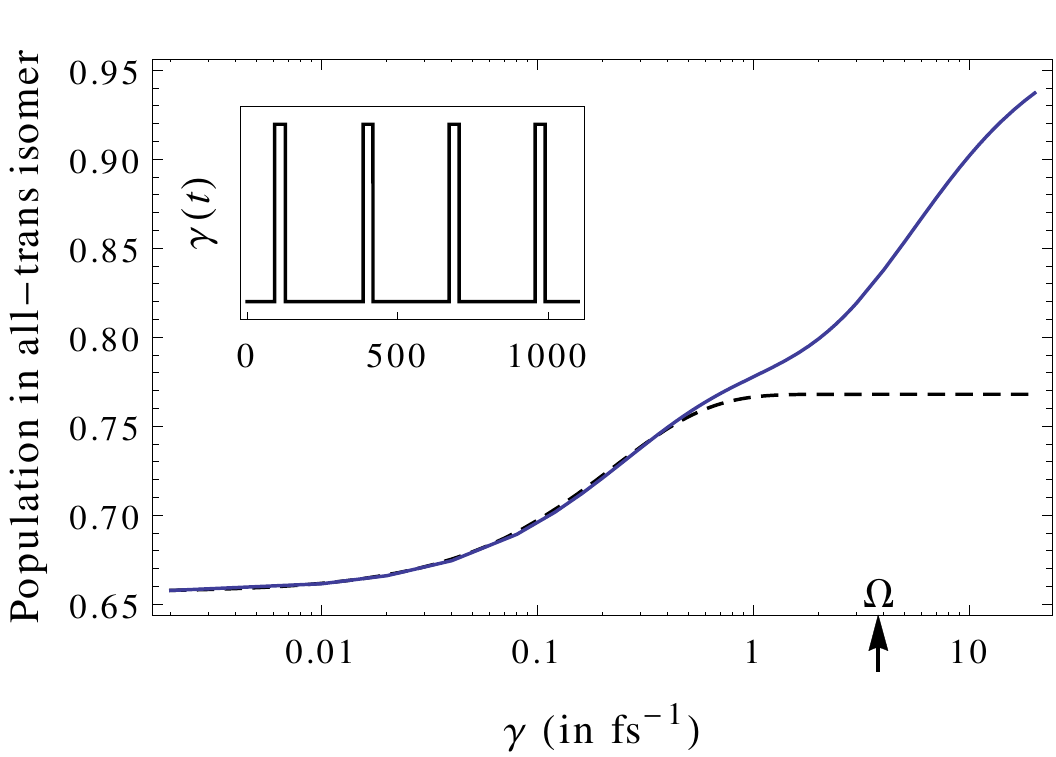}
  \caption{Final isomerization yield for a pulsed nonselective measurement with rate $\gamma$ (solid line). For comparison, the dashed line shows the result of the analytical open Landau-Zener model. The inset indicates the periods when the measurement is switched on (abscissa in fs), corresponding to times when the excited state wave packet transits the avoided crossing from left to right. An all-trans population of 80\% is obtained already at a moderate measurement rate of $2\,\text{fs}^{-1}$. \label{fig:popg4}}
\end{figure}

We can now determine the {final} yield of the  full isomerization reaction when adequately timed nonselective measurements are applied. This is done by propagating Eq.~\eqref{isons} for the entire  isomerization  time $t = 1.1 \,\text{ps}$, with the measurements switched on during the four left-to-right transits of the avoided crossing, at the intervals $[90 \,\text{fs}, 120 \,\text{fs}]$, $[390 \,\text{fs}, 420 \,\text{fs}]$, $[670 \,\text{fs}, 700 \,\text{fs}]$, and $[960 \,\text{fs}, 990 \,\text{fs}]$. The solid line Fig.~\ref{fig:popg4} shows the resulting final all-trans population with the dashed line representing again the open Landau-Zener model for reference. The inset indicates the time dependent switching of the measurement laser (we chose rectangular pulses for definiteness---smoother pulse shapes, \eg due to timing errors, have no qualitative influence). The final isomerization yield increases monotonically with $\gamma$, starting from its experimentally observed value of 65\% at $\gamma = 0$, and eventually approaching unity. Comparing this to the all-trans population after a single transit in Fig.~\ref{fig:popg1}, one observes that the measurement rate required for achieving a given population is roughly one order of magnitude lower. A rate of $2\,\text{fs}^{-1}$ leads to a yield of 80\%.

In conclusion, we have seen that one can enhance the isomerization yield of retinal in the visual pigment protein rhodopsin by a continuous or pulsed excitation of an infrared or two-photon Raman transition at a frequency between 800 and $950 \,\text{cm}^{- 1}$. This scheme relies on the controlled measurement back-action arising from the fact that the ground and first excited electronic state are distinguished by their different infrared spectra. The expected enhancement surpasses what could be achieved in optimal coherent control schemes \cite{Miller2006} and the necessary optical controls are implemented more easily. The scheme is also more robust, since an intricately sculptured coherent control pulse loses its efficiency when subject to environmental noise, while the underlying decoherence effect is not prone to such errors. The switching of infrared radiation distinguishing specific molecular states seems therefore suitable for manipulating large molecules under ambient conditions where conventional control handles are unsuited.

\end{document}